\documentclass{elsart}

\usepackage{epsfig}

\usepackage{amssymb}

\begin{document}

\newcommand{\XI}{\ensuremath{\Xi^{-}}}
\newcommand{\XB}{\ensuremath{\overline{\Xi}}\ensuremath{^{+}}}

\begin{frontmatter}


\title{$\Xi^{-}$ and $\overline{\Xi}$$^{+}$ Production in Central Pb+Pb Collisions at 158 GeV/$c$ per Nucleon}

\noindent
\address{
S.V.~Afanasiev$^{9}$,T.~Anticic$^{20}$, D.~Barna$^{5}$,
J.~Bartke$^{7}$, R.A.~Barton$^{3}$,
L.~Betev$^{10}$, H.~Bia{\l}\-kowska$^{17}$, A.~Billmeier$^{10}$,
C.~Blume$^{8}$, C.O.~Blyth$^{3}$, B.~Boimska$^{17}$, M.~Botje$^{1}$,
J.~Bracinik$^{4}$, R.~Bramm$^{10}$, R.~Brun$^{11}$,
P.~Bun\v{c}i\'{c}$^{10,11}$, V.~Cerny$^{4}$,
J.G.~Cramer$^{19}$, P.~Csat\'{o}$^{5}$, P.~Dinkelaker$^{10}$,
V.~Eckardt$^{16}$, P.~Filip$^{16}$,
H.G.~Fischer$^{11}$, Z.~Fodor$^{5}$, P.~Foka$^{8}$, P.~Freund$^{16}$,
V.~Friese$^{15}$, J.~G\'{a}l$^{5}$,
M.~Ga\'zdzicki$^{10}$, G.~Georgopoulos$^{2}$, E.~G{\l}adysz$^{7}$, 
S.~Hegyi$^{5}$, C.~H\"{o}hne$^{15}$, G.~Igo$^{14}$,
P.G.~Jones$^{3}$, K.~Kadija$^{11,20}$, A.~Karev$^{16}$,
V.I.~Kolesnikov$^{9}$, T.~Kollegger$^{10}$, M.~Kowalski$^{7}$, 
I.~Kraus$^{8}$, M.~Kreps$^{4}$, M.~van~Leeuwen$^{1}$, 
P.~L\'{e}vai$^{5}$, A.I.~Malakhov$^{9}$, S.~Margetis$^{13}$,
C.~Markert$^{8}$, B.W.~Mayes$^{12}$, G.L.~Melkumov$^{9}$,
A.~Mischke$^{8}$,J.~Moln\'{a}r$^{5}$, J.M.~Nelson$^{3}$,
G.~P\'{a}lla$^{5}$, A.D.~Panagiotou$^{2}$,
K.~Perl$^{18}$, A.~Petridis$^{2}$, M.~Pikna$^{4}$, L.~Pinsky$^{12}$,
F.~P\"{u}hlhofer$^{15}$,
J.G.~Reid$^{19}$, R.~Renfordt$^{10}$, W.~Retyk$^{18}$,
C.~Roland$^{6}$, G.~Roland$^{6}$, A.~Rybicki$^{7}$, T.~Sammer$^{16}$,
A.~Sandoval$^{8}$, H.~Sann$^{8}$, N.~Schmitz$^{16}$, P.~Seyboth$^{16}$,
F.~Sikl\'{e}r$^{5}$, B.~Sitar$^{4}$, E.~Skrzypczak$^{18}$,
G.T.A.~Squier$^{3}$, R.~Stock$^{10}$, H.~Str\"{o}bele$^{10}$, T.~Susa$^{20}$,
I.~Szentp\'{e}tery$^{5}$, J.~Sziklai$^{5}$,
T.A.~Trainor$^{19}$, D.~Varga$^{5}$, M.~Vassiliou$^{2}$,
G.I.~Veres$^{5}$, G.~Vesztergombi$^{5}$,
D.~Vrani\'{c}$^{8}$, S.~Wenig$^{11}$, A.~Wetzler$^{10}$, C.~Whitten$^{14}$,
I.K.~Yoo$^{15}$, J.~Zaranek$^{10}$, J.~Zim\'{a}nyi$^{5}$ \\
$^{1}$NIKHEF, Amsterdam, Netherlands. \\
$^{2}$Department of Physics, University of Athens, Athens, Greece.\\
$^{3}$Birmingham University, Birmingham, England.\\
$^{4}$Comenius University, Bratislava, Slovakia.\\
$^{5}$KFKI Research Institute for Particle and Nuclear Physics, Budapest, Hungary.\\
$^{6}$MIT, Cambridge, USA.\\
$^{7}$Institute of Nuclear Physics, Cracow, Poland.\\
$^{8}$Gesellschaft f\"{u}r Schwerionenforschung (GSI), Darmstadt, Germany.\\
$^{9}$Joint Institute for Nuclear Research, Dubna, Russia.\\
$^{10}$Fachbereich Physik der Universit\"{a}t, Frankfurt, Germany.\\
$^{11}$CERN, Geneva, Switzerland.\\
$^{12}$University of Houston, Houston, TX, USA.\\
$^{13}$Kent State University, Kent, OH, USA.\\
$^{14}$University of California at Los Angeles, Los Angeles, USA.\\
$^{15}$Fachbereich Physik der Universit\"{a}t, Marburg, Germany.\\
$^{16}$Max-Planck-Institut f\"{u}r Physik, Munich, Germany.\\
$^{17}$Institute for Nuclear Studies, Warsaw, Poland.\\
$^{18}$Institute for Experimental Physics, University of Warsaw, Warsaw, Poland.\\
$^{19}$Nuclear Physics Laboratory, University of Washington, Seattle, WA, USA.\\
$^{20}$Rudjer Boskovic Institute, Zagreb, Croatia.\\
}

\begin{abstract}

Results of the production of \XI~and \XB~hyperons in central Pb+Pb interactions at 158 GeV/$c$ per nucleon are presented.  This analysis utilises a global reconstruction procedure, which allows a measurement of 4$\pi$ integrated yields to be made for the first time.  Inverse slope parameters, which are determined from an exponential fit to the transverse mass spectra, are found to be 267 $\pm$ 9 MeV and 296 $\pm$ 17 MeV for \XI~and \XB~hyperons respectively.  Central rapidity densities (d$N$/d$y$) are found to be 1.49 $\pm$ 0.08 and 0.33 $\pm$ 0.04 particles per event per unit rapidity for \XI~and \XB~respectively.  The ratio $R$(\XB/\XI) at midrapidity, $R$($y=y_{\rm{cm}}$), is found to be 0.22 $\pm$ 0.03.  Yields integrated to full phase space are found to be 4.12 $\pm$ 0.20 and 0.77 $\pm$ 0.04 particles per event for \XI~and \XB~hyperons respectively.

\end{abstract}

\end{frontmatter}


\section{Introduction}

High energy heavy--ion physics provides a unique opportunity to study nuclear matter under extreme conditions of temperature and pressure.  Quantum Chromodynamics (QCD) predicts that at high energy density, nuclear matter will melt into a deconfined state of partonic matter known as the Quark--Gluon Plasma (QGP).  Experimental efforts to detect this novel phase rely on measurements of final state particles emitted from the fireball created by colliding nuclear matter.  There are three key stages in the collision process in which a QGP is formed: the initial partonic interactions which occur as two nuclei interpenetrate, the subsequent partonic interactions in a thermalised QGP and finally hadronic interactions once the temperature of the system has dropped below the critical temperature.  In order to probe the QGP, signals must be sought which are produced dominantly in the equilibrated deconfined state and survive the hadronisation process.   (For a general overview, we refer to the proceedings of the Quark Matter '99 conference \cite{qm99}.)

For almost 20 years, the production of strangeness has been proposed as a sensitive signal to QGP formation \cite{raforig}.  (For a recent overview see the proceedings of the Strangeness 2000 conference \cite{sqm2000}.)  Strange particles are of particular interest in hadronic collisions since they carry a new quantum number not present in the colliding nucleons or nuclei. In a QGP, $s\overline{s}$ pairs are readily created because of the newly available gluonic degrees of freedom which contribute approximately 80$\%$ to the total strangeness produced \cite{rafkock}. Strangeness in a partonic system where chiral symmetry is partially restored, is also expected to be readily produced due to the lower energy threshold for $s\overline{s}$ quark pair production.  At finite baryon density, where the chemical potential associated with the production of further light quarks is raised, the production of strangeness may be further increased.  These considerations lead to the expectation that the QGP phase should be characterised by a high strange (anti)quark density \cite{raf}.  By virtue of their strange quark content, multi--strange baryons and anti--baryons are expected to be a more sensitive probe to this phase than hadrons which contain only one strange valence quark.   The production of multi--strange baryons, and particularly multi--strange anti--baryons is otherwise suppressed by high hadronic energy thresholds as well as by long timescales for multi--step processes.  The NA49 \cite{frankxi} and WA97 \cite{wa97_enh} collaborations have previously reported an enhancement of multi--strange hyperons from central Pb+Pb collisions at 158 GeV/$c$ per nucleon at midrapidity compared to a convolution of nucleon-nucleon collisions.  Total yields of strange hadrons provide additional information, since they are an essential ingredient in thermal models \cite{cley_4pi}.  Remarkably, these models have described a wide range of experimental data with much success \cite{therm_success,eepp_success,braun}. Notable exceptions are in the multi--strange sector \cite{antin}. This paper reports on the rapidity distributions of $\Xi^-$ and $\overline{\Xi}$$^{+}$ measured 
by the NA49 collaboration over a wider range than has previously been published.

\section{The NA49 Experiment and Global Analysis}

The NA49 large acceptance hadron spectrometer has been used to record the charged final state particles produced in central Pb+Pb collisions at 158 GeV per nucleon at the CERN SPS.    The main tracking devices are four large volume time projection chambers (TPC) which are capable of detecting 80$\%$ of some 1500 charged particles created in a central Pb+Pb collision.  Two TPCs, VT1 and VT2, are located in strong magnetic fields of 1.5 and 1.1 T respectively, which allows charge and momentum to be assigned to the detected particles.  A further two TPCs are situated downstream from the magnets in a field free region.  The results presented here were analysed with a global tracking scheme \cite{lsb_glob}, which combines track segments that belong to the same physical particle but are detected in different TPCs.  Generally, global tracks have better momentum resolution.   Utilising a global tracking scheme increases detector acceptance and improves the track reconstruction efficiency in the TPCs, because tracks measured in the lower density regions far from the interaction vertex can be used to identify track segments in high density regions. Full details of the NA49 detector set--up and performance of tracking are available in \cite{na49nim}.

The results contained in this paper were obtained from the analysis of 400,000 central Pb+Pb events. The events were selected according to the energy deposited in a forward calorimeter and correspond to the 10$\%$ most central collisions with an average number of participants $N_{part}$ $\approx$ 335.

\section{$\Xi^-$ and $\overline{\Xi}$$^+$ from Central Pb+Pb Collisions}
\label{ana}

\begin{figure}[h]
\begin{center}
\epsfig{figure=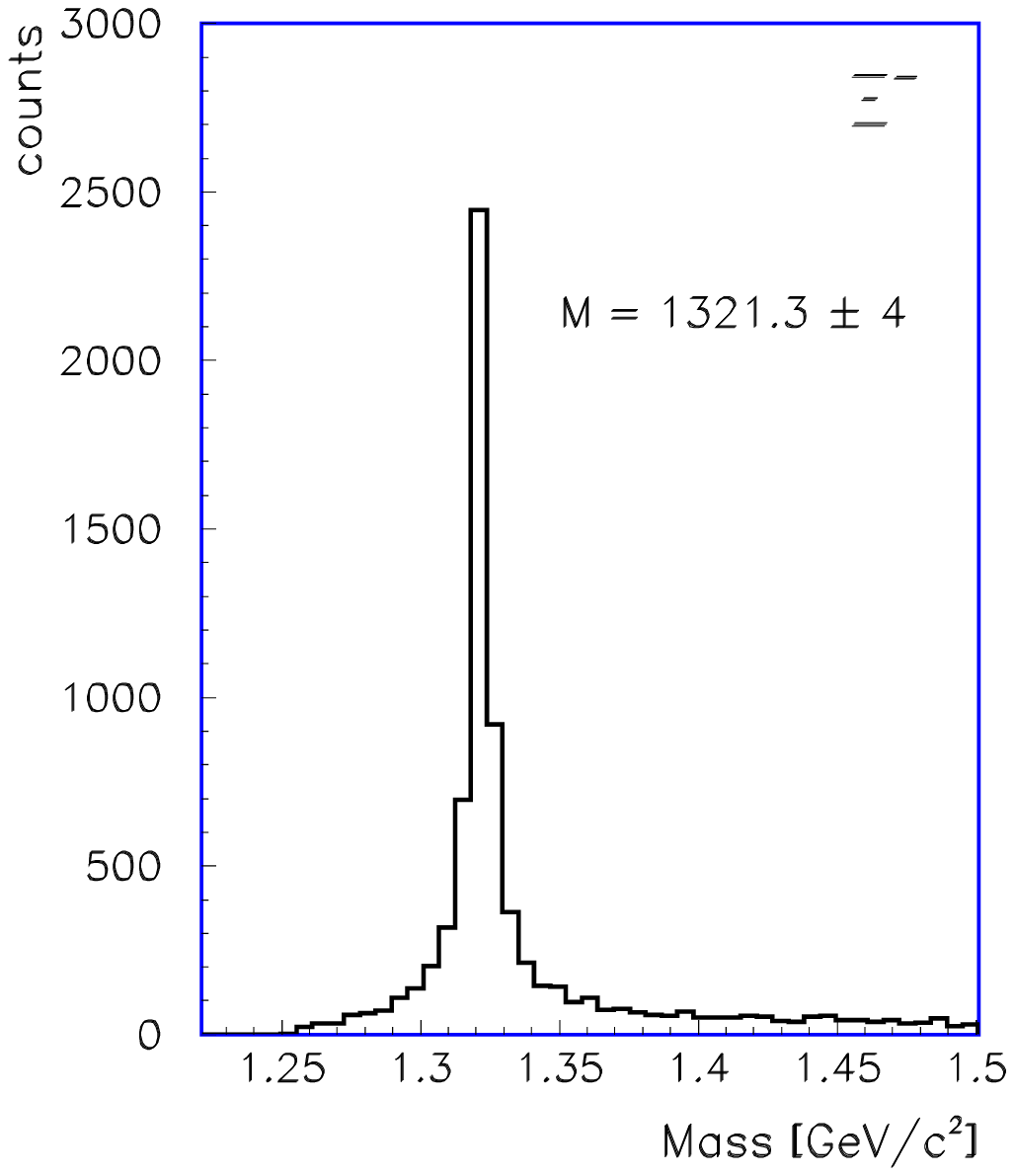,height=5.2cm,width=5.2cm}
\epsfig{figure=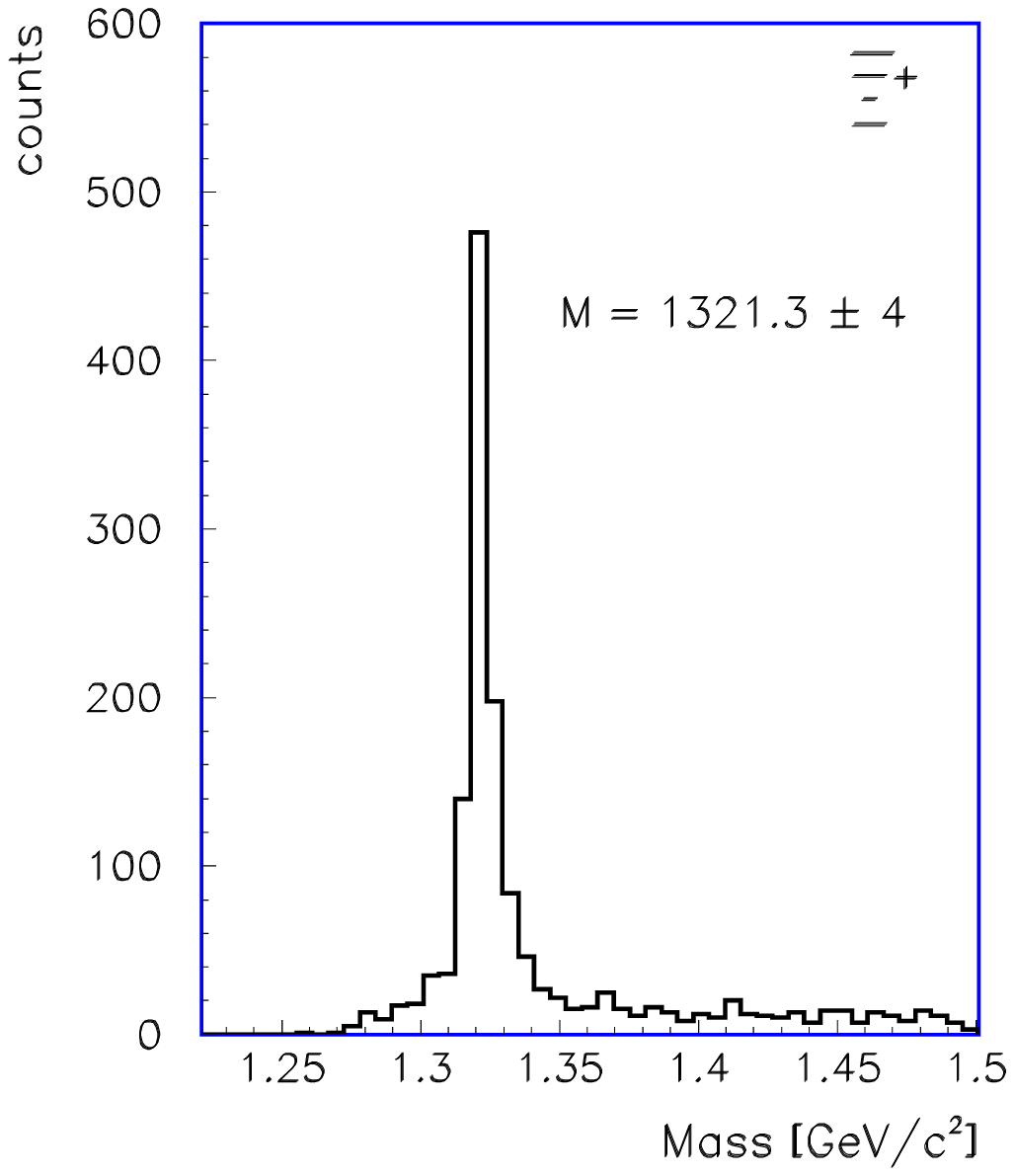,height=5.2cm,width=5.2cm}
\end{center}
\caption{Invariant mass distributions for $\Xi^-$ (left) and $\overline{\Xi}$$^+$ (right) from $400\,000$ central Pb+Pb collisions.  In the bin centred on the $\Xi^-$ mass, the signal--to--noise ratio is better than 10:1.}
\label{ximass}
\end{figure}

The doubly--strange $\Xi^-$ hyperon decays via the channel $\Xi^- \to \Lambda + \pi^-$ ($100\%$ branching ratio) with the subsequent decay $\Lambda \to p + \pi^-$ ($63.9\%$ branching ratio).  Neutral particle decay candidates were formed by considering all possible pairs of
positively and negatively charged tracks reconstructed in the TPCs. Each pair of
tracks was extrapolated from their most upstream point back towards the target.
If the distance of closest approach between the two tracks was less than 1 cm at
any point along their trajectory the pair was saved as a candidate
$\Lambda$ ($\overline{\Lambda}$). A large fraction of the combinatorial background arising from random pairs of tracks was eliminated by requiring that the tracks were separated by at 
least 2 cm at the target plane and that the decay vertex was at least 100 cm downstream of the
target. If the invariant mass of the candidates fell within $\pm 15$ MeV/$c^2$ of the
$\Lambda$ ($\overline{\Lambda}$) mass, they were selected as candidate daughters of $\Xi^-$($\overline{\Xi}$$^{+}$)
decays. Each $\Lambda$ candidate was paired with each negatively charged track in
the event, and both were extrapolated back toward the target. As in the case of
reconstructing the neutral decay vertices, if the distance of closest approach
between trajectories of the neutral $\Lambda$ candidate and the negatively charged
track fell within 1 cm, the pair was saved as a possible  $\Xi^-$ candidate.
Equivalently, each $\overline{\Lambda}$ candidate was paired with each positive track in
the event. Once again, straight-forward geometrical cuts were applied to eliminate
most of the combinatorial background. In this case, the most important cuts were
that the trajectories of the charged decay daughters were separated by between 2.5 cm and 7 cm from the primary vertex in the target plane depending on the position of the decay vertex.  An additional requirement was that the decay vertex was at least 35 cm downstream of the
target and that the $\Xi^-$($\overline{\Xi}$$^{+}$) candidate pointed back to the primary vertex. The
resulting invariant mass distributions of $\Xi^-$ and $\overline{\Xi}$$^{+}$ candidates are shown in
figure \ref{ximass}.

The signal was obtained by selecting those particles with invariant mass within $\pm 20$ MeV/$c^2$ of the $\Xi^-$ mass and then subtracting background counts from regions outside the peak.  Specifically, the background regions were defined between the ranges 1267 $< M <$ 1284 MeV/$c^2$ and 1358 $< M <$ 1381 MeV/$c^2$.  The background subtracted data set used consists of approximately 4800 $\Xi^-$ and 900 $\overline{\Xi}$$^+$ reconstructed particles \cite{rab_thesis}.

Those  $\Xi^-$ and $\overline{\Xi}$$^{+}$ which were excluded by the geometric selection
criteria, or not found due to reconstruction losses, were later accounted for by
applying a correction for the overall reconstruction inefficiency.

\subsection{Transverse Mass Distributions}

\begin{figure}[h]
\begin{center}
\epsfig{figure=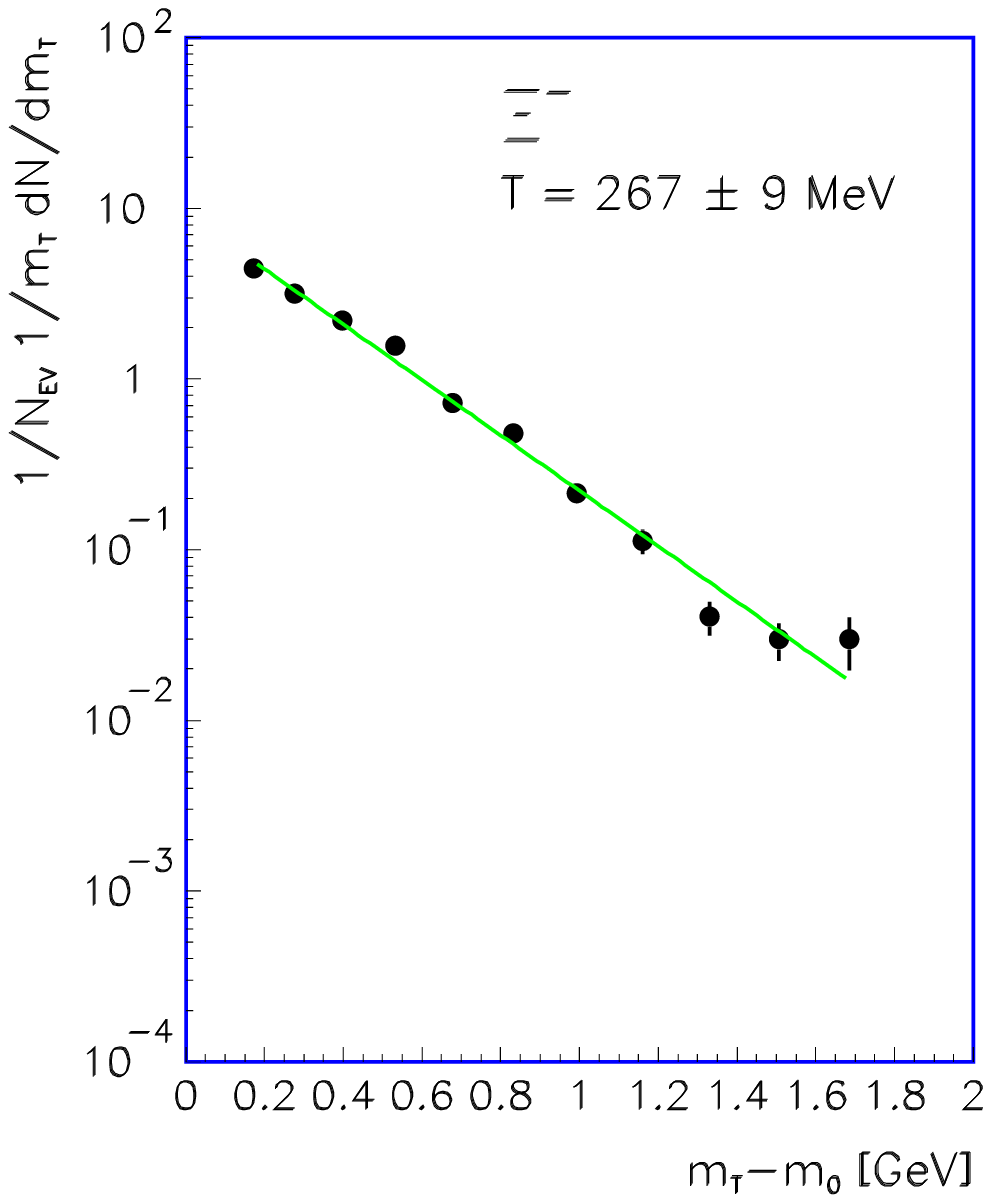, height=5.6cm, width=5.2cm}
\epsfig{figure=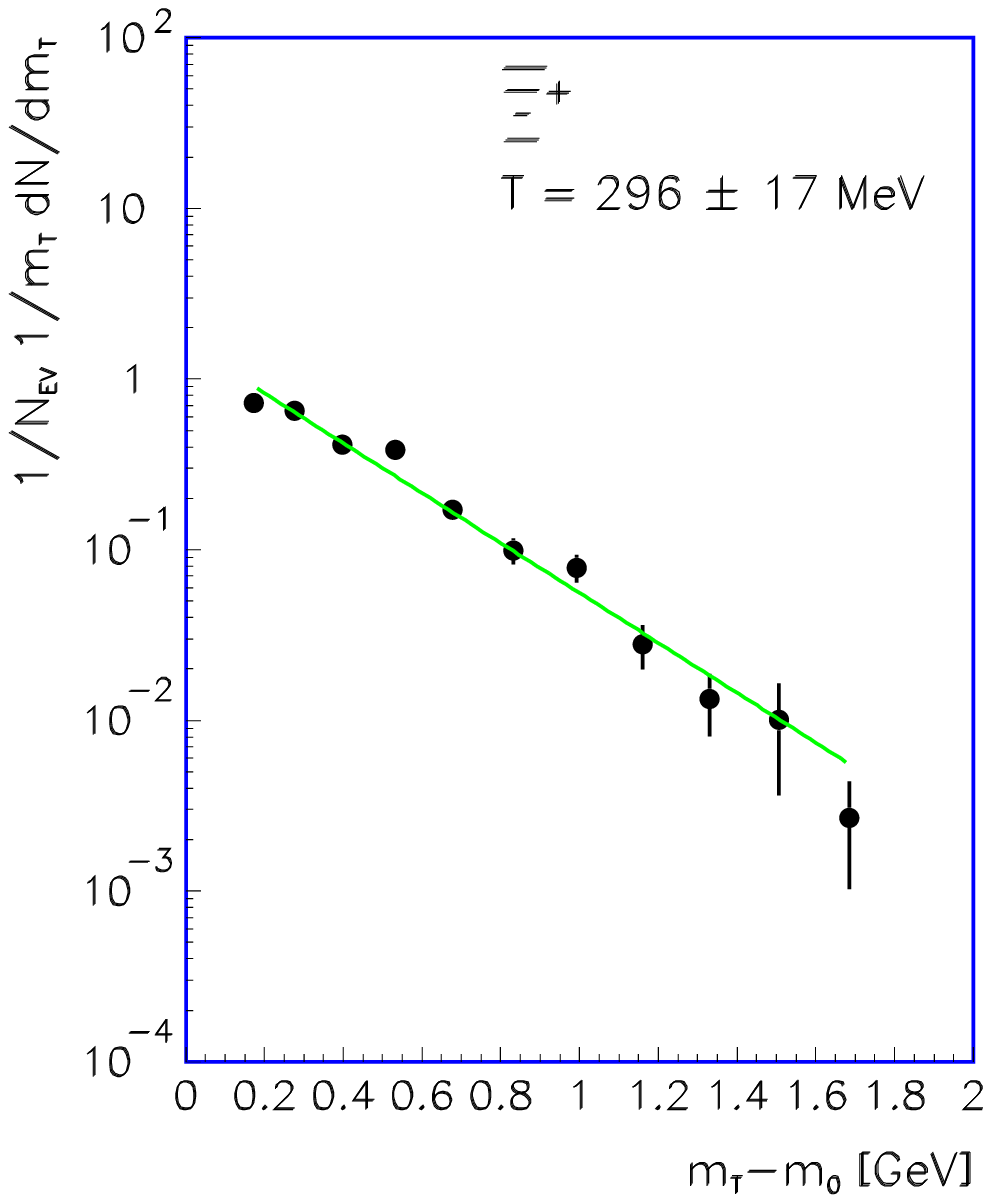, height=5.6cm, width=5.2cm}
\end{center}
\caption{Transverse mass distributions for $\Xi^-$ (left) and $\overline{\Xi}$$^+$ (right) from central Pb+Pb collisions.  Inverse slope parameters, $T$, are from fits to equation (\ref{mtfit}).}
\label{ximt}
\end{figure}

The data were evaluated in bins of rapidity, transverse momentum and lifetime.  Corrections for geometrical acceptance, branching ratio and reconstruction efficiency were then applied on a bin--by--bin basis.  The GEANT simulation package has been adapted for NA49 and was used for calculating the geometrical acceptance.  For the determination of the reconstruction efficiency, a single Monte Carlo $\Xi^-$ (or $\overline{\Xi}$$^+$) generated within the spectrometer acceptance was embedded into the raw data structures in which no real $\Xi^-$ (or $\overline{\Xi}$$^+$) was detected.  These events were processed with the reconstruction software which tried to reconstruct the embedded particle. The reconstruction efficiency of the embedded $\Xi^-$ (or $\overline{\Xi}$$^+$) was determined statistically for each bin of rapidity, transverse momentum and lifetime.  The kinematic phase space populated by detectable $\Xi^-$ and $\overline{\Xi}$$^+$ hyperons was found to be $0.6 < p_{T} < 2.8$ GeV/$c$ in transverse momentum and $1.7 < y < 4.5$ in rapidity.  Note that the rapidity range covers both sides of midrapidity ($y_{\rm{cm}} = 2.9$).  Fully corrected transverse mass ($m_{T} = \sqrt{{p_T}^2 + {m_0}^2}$) distributions are shown in figure \ref{ximt} and are fitted with the function given in equation (\ref{mtfit}) to obtain the inverse slope parameters, $T$. 

\begin{equation}
\frac{1}{m_T}\frac{dN}{dm_T} \propto \exp \left( \frac{-m_T}{T} \right)
\label{mtfit}
\end{equation}

The measured transverse mass spectra span the detectable rapidity range in each $p_{T}$ bin.  The fitted temperature parameters are found to be 267 $\pm$ 9 MeV and 296 $\pm$ 17 MeV for \XI~and \XB~hyperons respectively.

\subsection{Rapidity Distributions, 4$\pi$ Yields and Particle Ratios}

\begin{figure}[h]
\begin{center}
\epsfig{figure=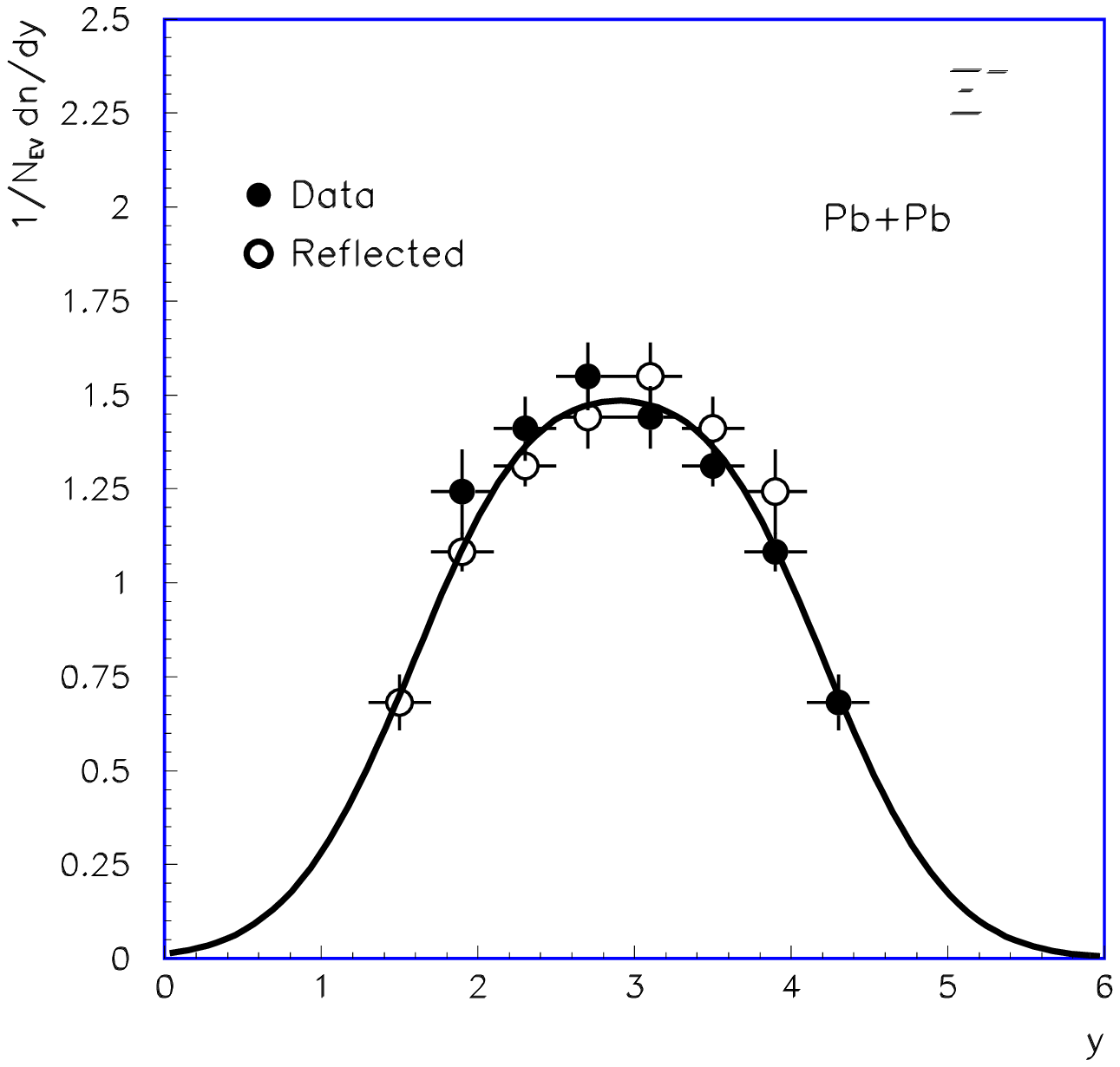,height=4.60cm,width=4.5cm}
\epsfig{figure=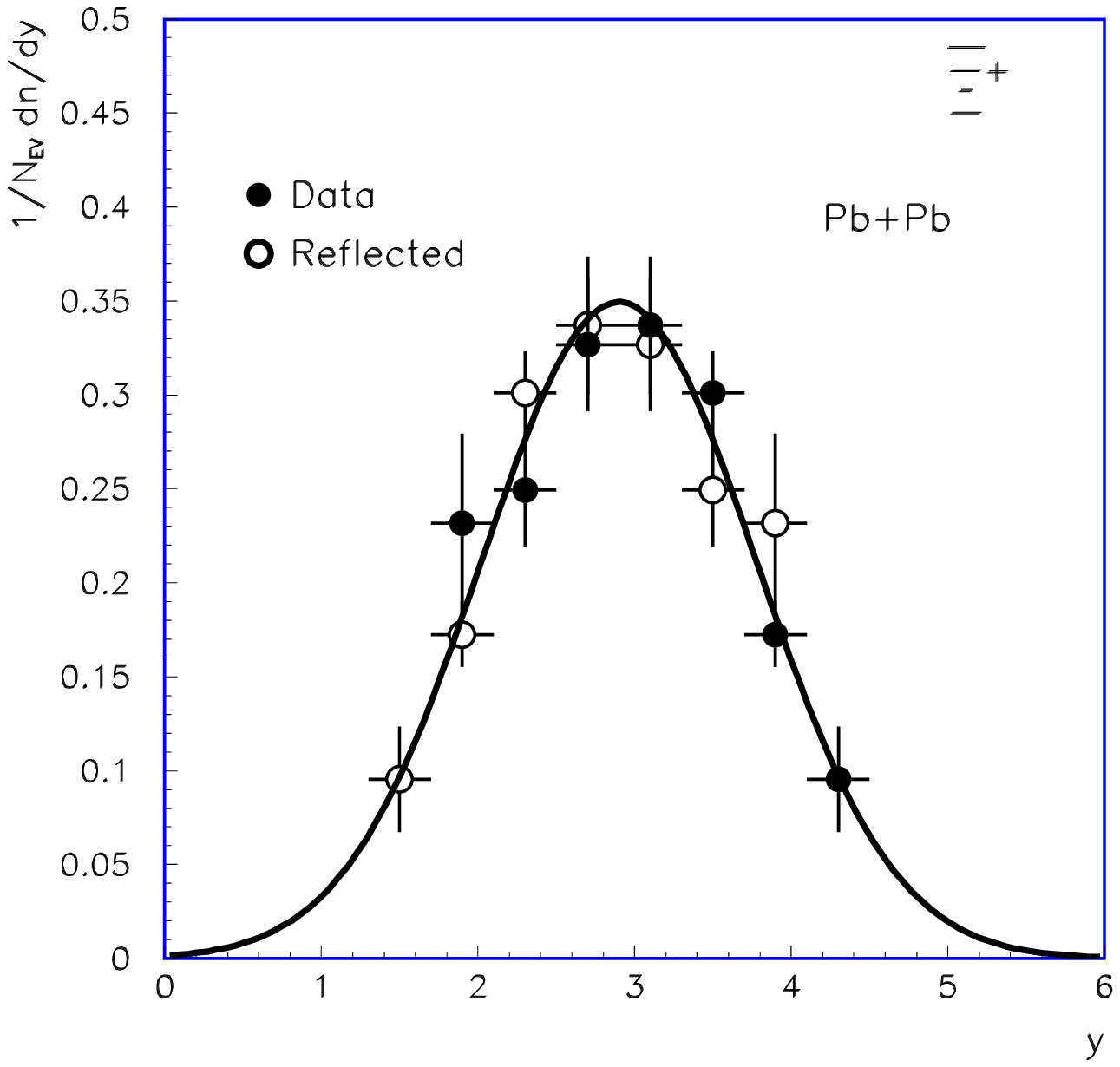,height=4.60cm,width=4.5cm}
\epsfig{figure=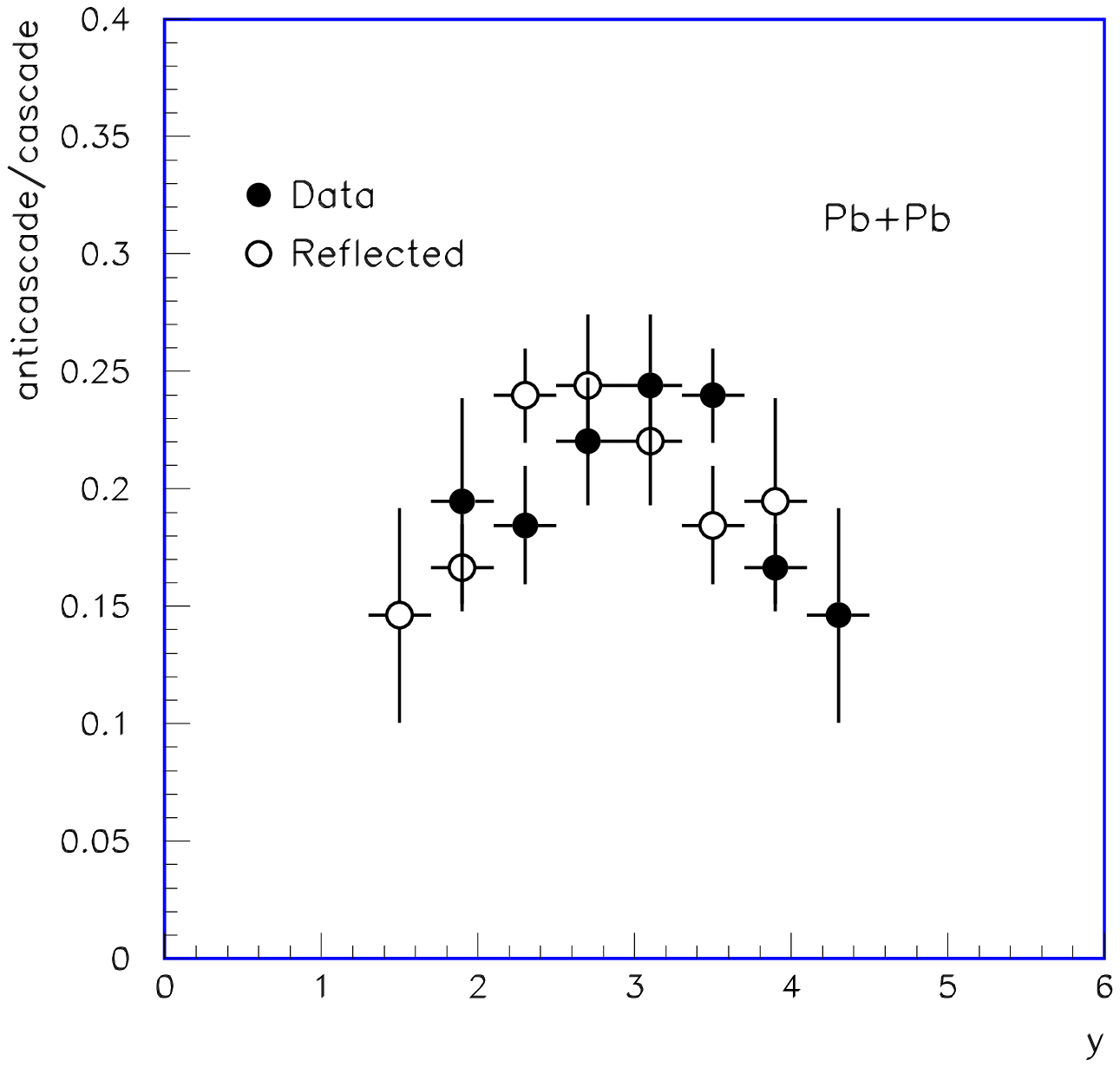,height=4.60cm,width=4.5cm}
\end{center}
\caption{Rapidity distributions for $\Xi^-$ (left), $\overline{\Xi}$$^+$ (centre) and the $\overline{\Xi}$$^+/\Xi^-$ ratio (right) from Pb+Pb collisions.  Closed circles are measured data points and open circles are reflected about midrapidity ($y$=2.9). The extrapolation factors are 1.51 and 1.45 for the  $\Xi^-$ and $\overline{\Xi}$$^+$ distributions, respectively.  The solid line fits are discussed in the text.}
\label{xirap}
\end{figure}

The large acceptance coverage of the NA49 spectrometer allows the experimental determination of fully integrated particle yields.  However, physical measurements are not possible for all values of transverse momentum because of finite geometrical acceptance.  Consequently, the rapidity spectra for $\Xi^-$ and $\overline{\Xi}$$^+$ are formed by extrapolating the transverse momentum distributions over the full range by means of eq. (1) using the fitted slope parameters, and assuming factorisation in $p_T$ and $y$.  The resulting distributions are shown in figure \ref{xirap}.  The extrapolation factors are 1.51 and 1.45 for the  $\Xi^-$ and $\overline{\Xi}$$^+$ distributions, respectively.  At midrapidity, the yields per event per unit of rapidity (or central rapidity densities, $dN/dy$) are found to be $1.49 \pm 0.08$ and $0.33 \pm 0.04$ for $\Xi^-$ and $\overline{\Xi}$$^+$ hyperons respectively. 

Integration of the measured part of the rapidity distribution and taking into account the reflected data, averaging where necessary, provides a lower limit for the cascade multiplicities of $3.76 \pm 0.12$ and $0.72 \pm 0.05$ particles per event for $\Xi^-$ and $\overline{\Xi}$$^+$ respectively. In order to estimate the 4$\pi$ yield, a fit to a double gaussian has been made in the case of the $\Xi^-$ rapidity spectrum. Each gaussian has a width $\sigma = 0.75 \pm 0.12$ centred at $y_{cm} \pm 0.66$.  The integral of this fit over the full rapidity range provides an estimated 4$\pi$ yield of $4.12 \pm 0.20$ $\Xi^-$ per event.  A single Gaussian centred at midrapidity with a width of $\sigma = 0.87 \pm 0.07$ is shown in figure \ref{xirap} for the $\overline{\Xi}$$^+$ distribution.  Integrating this fit over the full rapidity range gives an estimated 4$\pi$ yield of $0.77 \pm 0.04$ $\overline{\Xi}$$^+$ per event.  The $\overline{\Xi}$$^+ / \Xi^-$ ratio, $R$, can be calculated over the full measured rapidity range.  At midrapidity, $R$($y$=$y_{\rm cm}$) is found to be $0.22 \pm 0.03$ in good agreement with our previous publication \cite{frankxi} and with WA97 \cite{wa97xi}.  Note that in reference \cite{frankxi}, the event selection was based on the 5$\%$ most central collisions (10$\%$ in this paper), where $N_{part}$ was approximately 362 compared to 335 in this analysis.  Additionally, the large acceptance of the NA49 spectrometer allows a determination of the 4$\pi$ integrated ratio, which is found to be $R$(4$\pi$) $= 0.17 \pm 0.01$.

\subsection{Lifetime Distributions}

\begin{figure}[h]
\begin{center}
\epsfig{figure=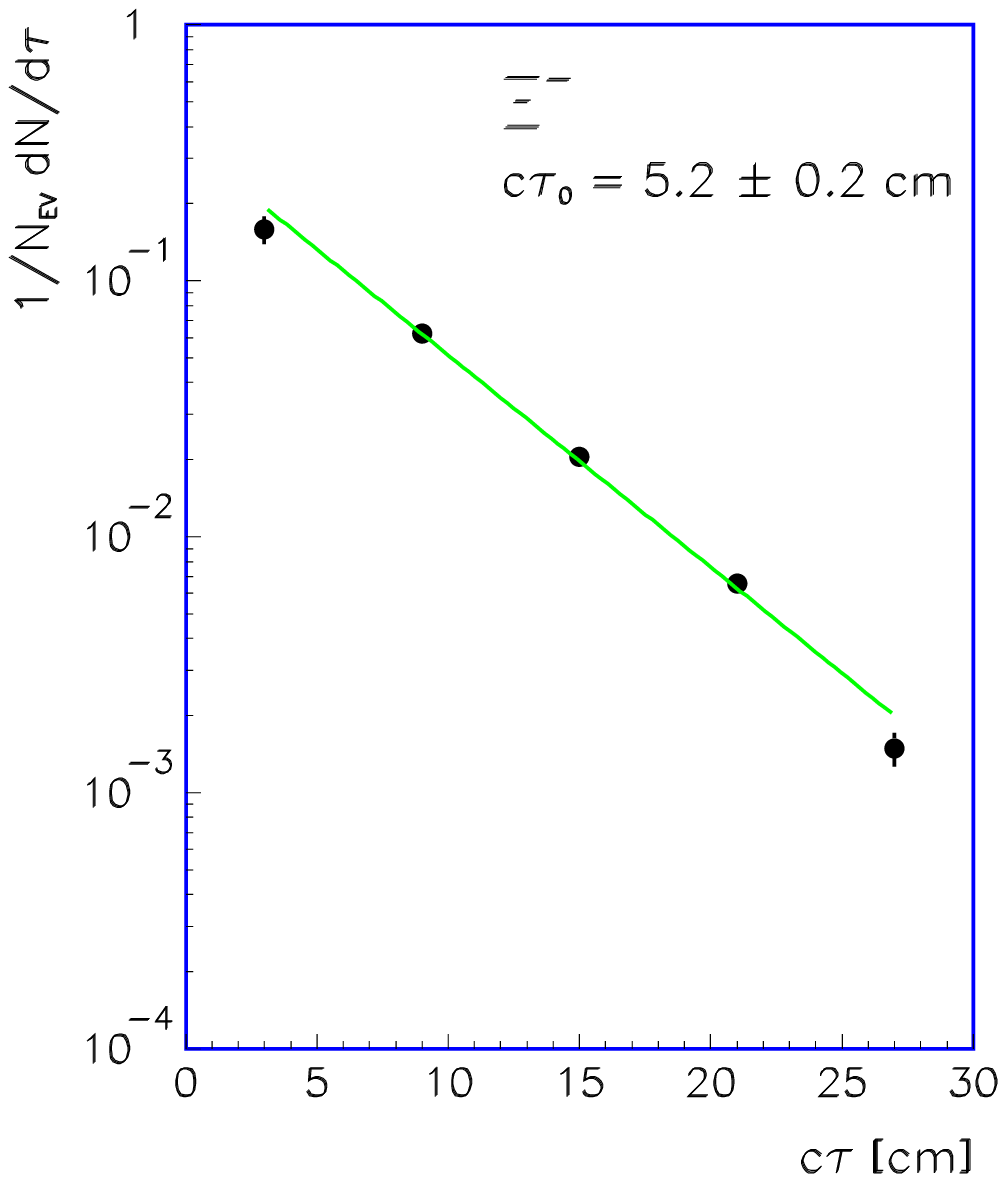,height=5.6cm,width=5.2cm}
\epsfig{figure=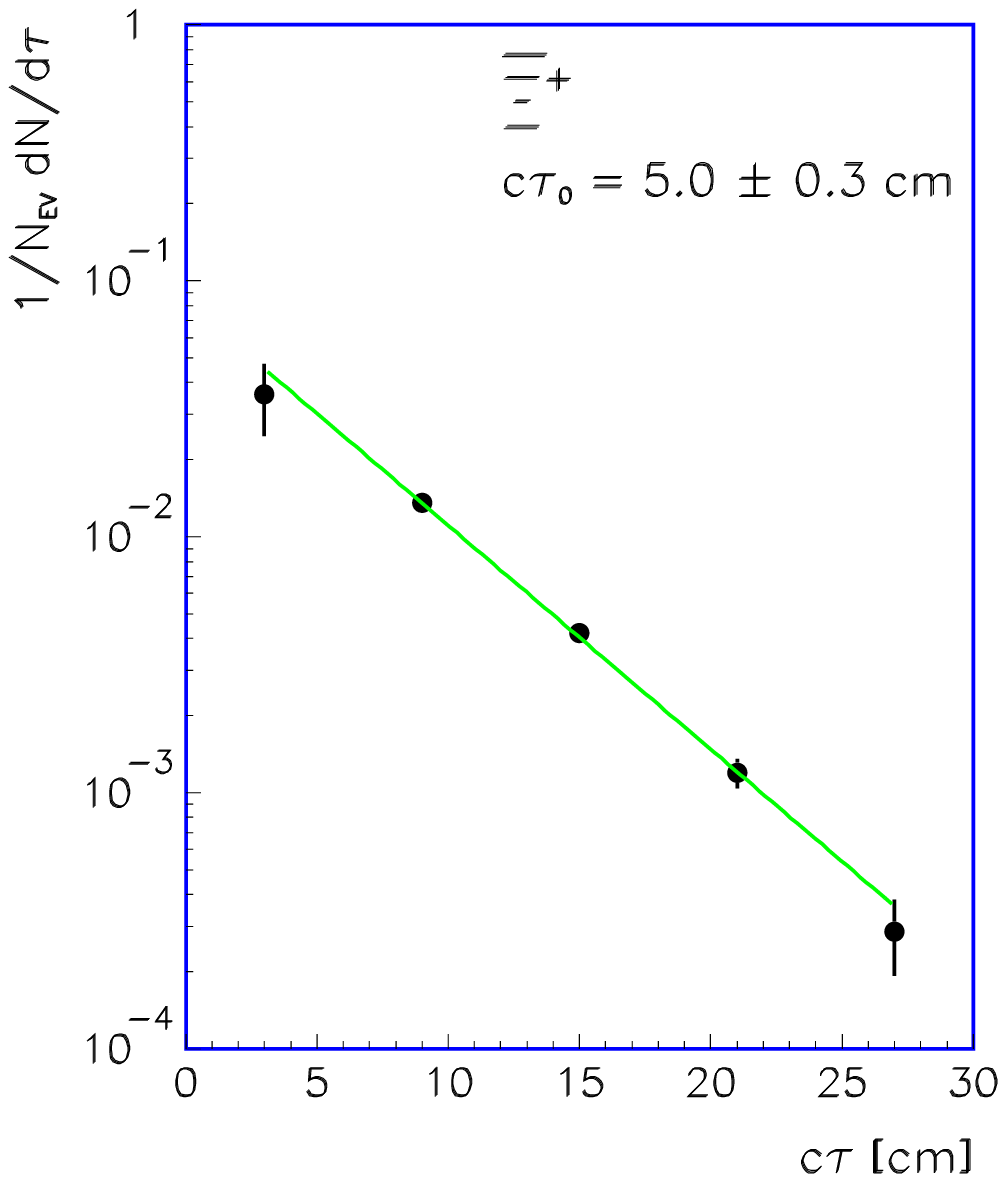,height=5.6cm,width=5.2cm}
\end{center}
\caption{Lifetime distributions for $\Xi^-$ (left) and $\overline{\Xi}$$^+$ (right) hyperons corrected for acceptance and efficiency.}
\label{xilife}
\end{figure}

In addition to dividing the data into $y$ and $p_T$ bins, the data were also evaluated in discrete bins of lifetime, $\tau$, allowing lifetime distributions to be measured.  Corrections for geometrical acceptance and reconstruction efficiency were applied in each of the lifetime bins and the resulting corrected distributions are shown in figure \ref{xilife}.  The distributions were fitted with exponentials to determine the mean lifetimes, which were found to be $5.2 \pm 0.2$ cm/$c$ and $5.0 \pm 0.3$ cm/$c$ for \XI~and \XB~hyperons respectively.  These values are consistent with the data book mean lifetime value of $\tau_{0} =$ 4.91 cm/$c$ \cite{data_book}.

\subsection{Discussion and Conclusions}

This new analysis uses a completely revised reconstruction procedure and covers a wide range in rapidity.  The resulting total integrated yields of $\Xi^-$ are 30$\%$ lower than previous NA49 estimates \cite{frankxi}.  This is due in part to the different centrality selection.  The current measurement benefits from a smaller error in extrapolation since we have been able to measure for the first time, the rapidity distributions of both $\Xi^-$ and $\overline{\Xi}$$^+$ over a wider range than previously. The central rapidity densities are in agreement with those found by WA97 \cite{pin_thesis}.  Errors on the new results are statistical only, with the systematic errors estimated to be about 15$\%$ derived from differences between simulation and data.

In summary, \XI~and \XB~hyperons have been measured at 158 GeV/$c$ per nucleon using the NA49 large acceptance spectrometer.  For the \XI~hyperons, the slope parameter determined from a fit to the transverse mass spectrum is 267 $\pm$ 9 MeV, the central rapidity density is found to be $1.49 \pm 0.08$ and the total yield is found to be $4.51 \pm 0.20$ particles per event integrated over all phase space.  The \XB~hyperons are found to have a slope parameter of 296 $\pm$ 17 MeV, with a central rapidity density of $0.33 \pm 0.04$ and a fully integrated yield of $0.77 \pm 0.04$ particles per event.  At midrapidity, the $\overline{\Xi}$$^+ / \Xi^-$ ratio is found to be $R$($y$=$y_{\rm cm}$) $= 0.22 \pm 0.03$.  A full phase space integrated $\overline{\Xi}$$^+ / \Xi^-$ ratio of $R$(4$\pi$) $= 0.19 \pm 0.01$ is obtained.

\subsection{Acknowledgements}

This work was supported by the Director, Office of Energy Research, Division of Nuclear Physics of the Office of High Energy and Nuclear Physics of the US Department of Energy (DE-ACO3-76SFOOO98 and DE-FG02-91ER40609), the US National Science Foundation, the Bundesministerium fur Bildung und Forschung, Germany, the Alexander von Humboldt Foundation, the UK Engineering and Physical Sciences Research Council, the Polish State Committee for Scientific Research (2 P03B 13820, 02615, 01716, and 09916), the Hungarian Scientific Research Foundation (T14920 and T23790), the EC Marie Curie Foundation, and the Polish-German Foundation.


 1

\end{document}